\documentstyle{article}

\begin{document}

\title{\flushright{\small KL-TH / 03-03} \bigskip \bigskip \bigskip \bigskip
\bigskip \bigskip \\
\center{ A new Map between Quantum Gauge Theories defined on a Quantum hyperplane
                                      and ordinary Gauge Theories: $q$-deformed QED}%
\thanks{%
Supported by DAAD}}
\author{L. Mesref\thanks{%
Email: lmesref@physik.uni-kl.de}}
\date{Department of Physics, Theoretical Physics\\
University of Kaiserslautern, Postfach 3049\\
67653 Kaiserslautern, Germany}
\maketitle

\begin{abstract}
We introduce a new map between a $q$-deformed gauge theory defined on a
general $GL_{q}\left( N\right) $-covariant quantum hyperplane and an
ordinary gauge theory \ in a full analogy with Seiberg-Witten map.
Perturbative analysis of the $q$-deformed QED at the classical level is
presented and gauge fixing \`{a} la BRST is discussed.

\newpage {}
\end{abstract}

\section{Introduction}

Motivated by the need to control the divergences in quantum electrodynamics,
Snyder \cite{snyder} proposed that one may use a noncommutative structure
for space-time coordinates. Although its great success, this suggestion has
been swiftly forsaken. This is partly due to a growing development in the
renormalization program which captivates all the attention of the leading
physicists. The renormalization prescription solves the quantum
inconsistencies without making any ad hoc assumptions on the space-time
structure. Thanks to the seminal paper of Connes \cite{connes1} the interest
in noncommutativity \cite{connes2} \ has been revived. Natural candidates
for noncommutativity are provided by quantum groups \cite{drinfeld}. A
special role is played by the quantum Yang-Baxter equations which express
the hidden symmetry of integrable systems . Nowadays many applications have
appeared. One can mention conformal field theories \cite{alvarez,moore},  as
well as in the vertex and spin models \cite{vega,pasquier}, , in quantum
optics \cite{chaichian} and quantum gauge theories \cite
{castellani,mesref1,mesref2}.

In a recent paper \cite{mesref3} we have constructed a new map which relates
a $q$-deformed gauge field defined on the Manin plane $\widehat{x}\widehat{y}%
=q\widehat{y}\widehat{x}$ and the ordinary gauge field. This map is the
analogue of the Seiberg-Witten map \cite{seiberg}. We have found this map
using the Gerstenhaber product \cite{gerstenhaber} instead of the
Groenewold-Moyal star product \cite{groenewold}. In the present letter we
extend our analysis to the general $GL_{q}\left( N\right) $-covariant
quantum hyperplane \cite{wess} generated by the coordinates $\widehat{x}%
^{1},...,\widehat{x}^{N}$ and defined by $\widehat{x}^{i}\widehat{x}%
^{j}=R_{kl}^{ij}\widehat{x}^{k}\widehat{x}^{l}$, where $R$ is the braiding
matrix.

This letter is organized as follows. In Sec. 2, We construct a new map
relating $q$-deformed and ordinary gauge fields. In Sec. 3, \ we present the
perturbative $q$-deformed QED at the classical level and we introduce the $q$%
-deformed BRST and anti-BRST transformations. This study prepare the
quantization of the model at hand.

\section{$q$-deformed gauge symmetry versus ordinary gauge symmetry}

The undeformed QED action is given by

\begin{equation}
S=\int d^{4}x\left[ \overline{\psi }\left( i\gamma ^{\mu }D_{\mu }-m\right)
\psi -\frac{1}{4}F_{\mu \nu }F^{\mu \nu }\right] ,  
\end{equation}

\bigskip

where

\bigskip

\begin{eqnarray}
D_{\mu }\psi &=&\left( \partial _{\mu }-iA_{\mu }\right) \psi ,  \nonumber \\
F_{\mu \nu } &=&\partial _{\mu }A_{\nu }-\partial _{\nu }A_{\mu }.
\end{eqnarray}

\bigskip

$S$ is invariant with respect to infinitesimal gauge transformations:

\bigskip

\begin{eqnarray}
\delta _{\alpha }A_{\mu } &=&\partial _{\mu }\alpha ,  \nonumber \\
\delta _{\alpha }\psi &=&i\alpha \psi ,  \nonumber \\
\delta _{\alpha }\overline{\psi } &=&-i\overline{\psi }\alpha . 
\end{eqnarray}

\bigskip

Let us now consider QED defined on a $GL_{q}\left( 4\right) $-covariant
quantum hyperplane $\widehat{x}^{i}\widehat{x}^{j}=q\widehat{x}^{j}\widehat{x%
}^{i},\quad i<j,\,\,q\in C.$

This relation is governed by the braiding $R$ matrix which is explicitly
given as:

\bigskip

\begin{equation}
R_{kl}^{ij}=\delta _{l}^{i}\delta _{k}^{j}\left( \left( 1-q^{-1}\right)
\delta ^{ij}+q^{-1}\right) +\left( 1-q^{-2}\right) \delta _{k}^{i}\delta
_{l}^{j}\Theta ^{ji}.  
\end{equation}

\bigskip

In the deformed case one replaces the ordinary product by t he Gerstenhaber
star product \cite{gerstenhaber} defined by

\bigskip

\begin{equation}
f\star g=\mu \circ e^{i\eta x^{i}\frac{\partial }{\partial x^{i}}\otimes
x^{j}\frac{\partial }{\partial x^{j}}}\left( f\otimes g\right) , 
\end{equation}

\bigskip

where the undeformed product $\mu $ is given by

\bigskip

\begin{equation}
\mu \left( f\otimes g\right) =fg. 
\end{equation}

\bigskip

A straightforward computation gives then the following commutation relations

\bigskip

\begin{equation}
x^{i}\star x^{j}=\sum_{r=0}\frac{\left( i\eta \right) ^{r}}{r!}%
x^{i}x^{j}=e^{i\eta }x^{i}x^{j},\quad x^{j}\star x^{i}=x^{j}x^{i},\quad
i<j\quad  
\end{equation}

\bigskip

whence

\bigskip

\begin{equation}
x^{i}\star x^{j}=qx^{j}\star x^{i}, \,\,\,\,\,\,\,\,q=e^{i\eta }.
\end{equation}

\bigskip

Thus we recover the commutation relations for the quantum hyperplane.

Let us illustrate by two examples:

The case $n=2$ (the Manin plane):

\bigskip

\begin{equation}
f\star g=\mu \circ e^{i\eta x\frac{\partial }{\partial x}\otimes y\frac{%
\partial }{\partial y}}\left( f\otimes g\right) . 
\end{equation}

\bigskip

Using this product it is not difficult to find the usual Manin plane
commutation relations: $x\star y=qy\star x.$

The case $n=3$, we have

\bigskip

\begin{equation}
f\star g=\mu \circ e^{i\eta \left( x\frac{\partial }{\partial x}\otimes y%
\frac{\partial }{\partial y}+x\frac{\partial }{\partial x}\otimes z\frac{%
\partial }{\partial z}+y\frac{\partial }{\partial y}\otimes z\frac{\partial 
}{\partial z}\right) }\left( f\otimes g\right) .
\end{equation}

\bigskip

We can easily prove the commutation relations: $x\star y=qy\star x,\quad
x\star z=qz\star x$ and $y\star z=qz\star x$.

If the spacetime dimension of the quantum hyperplane is $n$ we have $\frac{%
n\left( n-1\right) }{2}$ terms present in the tensor product.

Let us take $\theta ^{ij}\left( x\right) =\eta x^{i}x^{j}$ with $i<j$. Using
the expansion of (5) in $\eta $ we find

\bigskip

\begin{equation}
f\star g=fg+i\theta ^{ij}\left( x\right) \frac{\partial }{\partial x^{i}}f%
\frac{\partial }{\partial x^{j}}g+\circ \left( \eta ^{2}\right) , \quad i<j.
\end{equation}

\bigskip

The $q$-deformed infinitesimal gauge transformations are defined by

\begin{eqnarray}
\widehat{\delta }_{\widehat{\alpha }}\widehat{A}_{\mu } &=&\partial _{\mu }%
\widehat{\alpha }+i\left[ \widehat{\alpha },\widehat{A}_{\mu }\right]
_{\star }=\partial _{\mu }\widehat{\alpha }+i\widehat{\alpha }\star \widehat{%
A}_{\mu }-i\widehat{A}_{\mu }\star \widehat{\alpha },  \nonumber \\
\widehat{\delta }_{\widehat{\alpha }}\widehat{\psi } &=&i\widehat{\alpha }%
\star \widehat{\psi },  \nonumber \\
\widehat{\delta }_{\widehat{\alpha }}\widehat{\overline{\psi }} &=&-i%
\widehat{\overline{\psi }}\star \widehat{\alpha },  \nonumber \\
\widehat{\delta }_{\widehat{\alpha }}\widehat{F}_{\mu \nu } &=&i\widehat{%
\alpha }\star \widehat{F}_{\mu \nu }-i\widehat{F}_{\mu \nu }\star \widehat{%
\alpha }.  
\end{eqnarray}

\bigskip

To first order in $\eta $, the above formulas for the gauge transformations
read

\bigskip

\begin{eqnarray}
\widehat{\delta }_{\widehat{\alpha }}\widehat{A}_{\mu } &=&\partial _{\mu }%
\widehat{\alpha }-\theta ^{\rho \sigma }\left( x\right) \left( \partial
_{\rho }\alpha \partial _{\sigma }A_{\mu }-\partial _{\rho }A_{\mu }\partial
_{\sigma }\alpha \right) +o\left( \eta ^{2}\right) ,  \nonumber \\
\widehat{\delta }_{\widehat{\alpha }}\widehat{\psi } &=&i\widehat{\alpha }%
\widehat{\psi }-\theta ^{\rho \sigma }\left( x\right) \partial _{\rho
}\alpha \partial _{\sigma }\psi +o\left( \eta ^{2}\right) ,  \nonumber \\
\widehat{\delta }_{\widehat{\alpha }}\widehat{\overline{\psi }} &=&-i%
\widehat{\overline{\psi }}\widehat{\alpha }+\theta ^{\rho }\left( x\right)
\partial _{\rho }\overline{\psi }\partial _{\sigma }\alpha +o\left( \eta
^{2}\right) ,  \nonumber \\
\widehat{\delta }_{\widehat{\alpha }}\widehat{F}_{\mu \nu } &=&-\theta
^{\rho \sigma }\left( x\right) \left( \partial _{\rho }\alpha \partial
_{\sigma }F_{\mu \nu }-\partial _{\rho }F_{\mu \nu }\partial _{\sigma
}\alpha \right) +o\left( \eta ^{2}\right) . 
\end{eqnarray}

\bigskip

The solutions are given by

\bigskip

\begin{eqnarray}
\widehat{A}_{\mu } &=&A_{\mu }+\theta ^{\rho \sigma }\left( x\right) \left(
A_{\sigma }F_{\sigma \mu }-A_{\rho }\partial _{\sigma }A_{\mu }\right)
-\partial _{\mu }\theta ^{\rho \sigma }\left( x\right) A_{\sigma }A_{\rho
}+o\left( \eta ^{2}\right) ,  \nonumber \\
\widehat{\psi } &=&\psi -\theta ^{\rho \sigma }\left( x\right) A_{\rho
}\partial _{\sigma }\psi +o\left( \eta ^{2}\right) ,  \nonumber \\
\widehat{\alpha } &=&\alpha -\theta ^{\rho \sigma }\left( x\right) A_{\rho
}\partial _{\sigma }\alpha +o\left( \eta ^{2}\right) . 
\end{eqnarray}

\bigskip

An additional term proportional to $\partial _{\mu }\theta ^{\rho \sigma }\left( x\right) $ appears. In the case where $\theta ^{\rho \sigma }$ is real and antisymmetric we find exactly the Seiberg Witten map.

The $q$-deformed curvature $\widehat{F}_{\mu \nu }$ is given by

\bigskip

\begin{eqnarray}
\widehat{F}_{\mu \nu } &=&\partial _{\mu }\widehat{A}_{\nu }-\partial _{\nu }%
\widehat{A}_{\mu }-i\left[ \widehat{A}_{\mu },\widehat{A}_{\nu }\right]
_{\star }  \nonumber \\
&=&\partial _{\mu }\widehat{A}_{\nu }-\partial _{\nu }\widehat{A}_{\mu }-i%
\widehat{A}_{\mu }\star \widehat{A}_{\nu }+i\widehat{A}_{\nu }\star \widehat{%
A}_{\mu }.  
\end{eqnarray}

\bigskip

Using (11) and (14) we find

\bigskip

\begin{eqnarray}
\widehat{F}_{\mu \nu } &=&F_{\mu \nu }+\theta ^{\rho \sigma }\left( x\right)
\left( A_{\sigma }\partial _{\rho }F_{\mu \nu }-A_{\rho }\partial _{\sigma
}F_{\mu \nu }+F_{\rho \mu }F_{\sigma \nu }+F_{\mu \sigma }F_{\rho \nu
}\right)   \nonumber \\
&&+\partial _{\mu }\theta ^{\rho \sigma }\left( x\right) \left( A_{\sigma
}F_{\rho \nu }-A_{\rho }\partial _{\sigma }A_{\nu }\right)   \nonumber \\
&&-\partial _{\nu }\theta ^{\rho \sigma }\left( x\right) \left( A_{\sigma
}F_{\rho \mu }-A_{\rho }\partial _{\sigma }A_{\mu }\right) +o\left( \eta
^{2}\right) , 
\end{eqnarray}

\bigskip

which we can write as

\bigskip

\begin{equation}
\widehat{F}_{\mu \nu }=F_{\mu \nu }+f_{\mu \nu }+o\left( \eta ^{2}\right) ,
\end{equation}

\bigskip

where $f_{\mu \nu }$ is the quantum correction linear in $\eta $. The
quantum analogue of the action (1) is given by

\bigskip

\begin{equation}
\widehat{S}=\int d^{4}x\quad \left[ \widehat{\overline{\psi }}\star \left(
i\gamma ^{\mu }\widehat{D}_{\mu }-m\right) \widehat{\psi }-\frac{1}{4}%
\widehat{F}_{\mu \nu }\star \widehat{F}^{\mu \nu }\right] .  
\end{equation}

\bigskip

Even for functions $f$, $g$ that vanish rapidly enough at infinity,we have%
\newline

\begin{equation}
\int Tr\,\,\widehat{f}\star \widehat{g}\neq \int Tr\,\,\widehat{f}\,\widehat{g}. 
\end{equation}

This situation is in contrast with noncommutative geometry where the
equality holds.

\section{Perturbative $q$-deformed QED and gauge fixing \`{a} la BRST}

The $q$-deformed action (18)

\bigskip

\begin{equation}
\widehat{S}=S+\widehat{S}_{q}  
\end{equation}

\bigskip

where $S$ is the undeformed action (1) and $\widehat{S}_{q}$ is the correction
linear in $\eta $ and given by

\bigskip

\begin{eqnarray}
\widehat{S}_{q} &=&\int d^{4}x\quad \theta ^{\rho \sigma }\left( x\right) \,%
{\LARGE [}\overline{\psi }\gamma ^{\mu }(A_{\sigma }F_{\rho \mu }-A_{\rho
}\partial _{\sigma }A_{\mu })\psi   \nonumber \\
&&-\overline{\psi }\gamma ^{\mu }A_{\mu }A_{\rho }\partial _{\sigma }\psi
-\partial _{\rho }\overline{\psi }\gamma ^{\mu }\partial _{\mu }\partial
_{\sigma }\psi -A_{\rho }\partial _{\sigma }\overline{\psi }\gamma ^{\mu
}A_{\mu }\psi   \nonumber \\
&&-mA_{\rho }\partial _{\sigma }\overline{\psi }\psi -m\overline{\psi }%
A_{\rho }\partial _{\sigma }\overline{\psi }-\frac{1}{2}(A_{\sigma }\partial
_{\rho }F_{\mu \nu }F^{\mu \nu }-  \nonumber \\
&&A_{\rho }\partial _{\sigma }F_{\mu \nu }F^{\mu \nu }+F_{\rho \mu
}F_{\sigma \nu }F^{\mu \nu }+F_{\mu \sigma }F_{\rho \nu }F^{\mu \nu })
\nonumber \\
&&-i{\LARGE (}\overline{\psi }\gamma ^{\mu }\partial _{\mu }\left( A_{\rho
}\partial _{\sigma }\psi \right) +A_{\rho }\partial _{\sigma }\overline{\psi 
}\gamma ^{\mu }\partial _{\mu }\psi   \nonumber \\
&&-\overline{\psi }\gamma ^{\mu }\partial _{\rho }A_{\mu }\partial _{\sigma
}\psi -\partial _{\rho }\overline{\psi }\partial _{\sigma }\left( A_{\mu
}\psi \right)   \nonumber \\
&&-m\partial _{\rho }\overline{\psi }\partial _{\sigma }\psi -\frac{1}{4}%
\partial _{\rho }F_{\mu \nu }\partial _{\sigma }F^{\mu \nu }{\LARGE )]} 
\nonumber \\
&&-\partial _{\mu }\theta ^{\rho \sigma }\left( x\right) {\LARGE [}\frac{1}{2%
}(A_{\sigma }F_{\rho \nu }F^{\mu \nu }-A_{\rho }\partial _{\sigma }A_{\nu
}F^{\mu \nu })  \nonumber \\
&&\qquad \qquad \qquad -i\,\gamma ^{\mu }A_{\rho }\partial _{\sigma }%
\overline{\psi }\psi {\LARGE ]}  \nonumber \\
&&+\frac{1}{2}\partial _{\upsilon }\theta ^{\rho \sigma }\left( x\right)
(A_{\sigma }F_{\rho \mu }F^{\mu \nu }-A_{\rho }\partial _{\sigma }A_{\mu
}F^{\mu \nu })+o\left( \eta ^{2}\right) .   
\end{eqnarray}

\bigskip

A gauge fixing term is needed in order to quantize the system. This is done
in the BRST and anti-BRST formalism. The quantum BRST transformations are
given by \cite{becchi}:

\bigskip

\begin{eqnarray}
\widehat{s}\widehat{A}_{\mu } &=&\partial _{\mu }\widehat{c}-\theta ^{\rho
\sigma }\left( x\right) \left( \partial _{\rho }c\partial _{\sigma }A_{\mu
}-\partial _{\rho }A_{\mu }\partial _{\sigma }c\right) +o\left( \eta
^{2}\right) ,  \nonumber \\
\widehat{s}\widehat{\psi } &=&i\widehat{c}\widehat{\psi }-\theta ^{\rho
\sigma }\left( x\right) \partial _{\rho }c\partial _{\sigma }\psi +o\left(
\eta ^{2}\right) ,  \nonumber \\
\widehat{s}\widehat{\overline{\psi }} &=&-i\widehat{\overline{\psi }}%
\widehat{c}+\theta ^{\rho }\left( x\right) \partial _{\rho }\overline{\psi }%
\partial _{\sigma }c+o\left( \eta ^{2}\right) ,  \nonumber \\
\widehat{s}\widehat{F}_{\mu \nu } &=&-\theta ^{\rho \sigma }\left( x\right)
\left( \partial _{\rho }c\partial _{\sigma }F_{\mu \nu }-\partial _{\rho
}F_{\mu \nu }\partial _{\sigma }c\right) +o\left( \eta ^{2}\right) , 
\nonumber \\
\widehat{s}\widehat{\overline{c}} &=&b,\quad \widehat{s}\widehat{c}=0,\quad 
\widehat{s}\widehat{b}=0, 
\end{eqnarray}

\bigskip

where $\widehat{c}$, $\widehat{\overline{c}}$ are the quantum Faddeev-Popov
ghost and anti-ghost fields, $\widehat{b}$ a scalar field (sometimes called
the Nielsen-Lautrup auxiliary field) and $\widehat{s}$ the quantum BRST
operator. The gauge-fixing term action is introduced as

\begin{equation}
\widehat{S}_{gf}=\int d^{4}x\quad \widehat{s}\left( \widehat{\overline{c}}%
\star \left( \frac{\alpha }{2}\widehat{b}-\partial _{\mu }\widehat{A}^{\mu
}\right) \right) \left( x\right) . 
\end{equation}

An expansion in $\eta $ leads to

\bigskip

\begin{eqnarray}
\widehat{S}_{gf} &=&\int d^{4}x\quad \frac{\alpha }{2}b^{2}+\overline{c}%
\partial ^{2}c-b\partial ^{\mu }A_{\mu }  \nonumber \\
&&-\theta ^{\rho \sigma }\left( x\right) {\LARGE [}b\partial ^{\mu }\left(
A_{\sigma }F_{\rho \mu }-A_{\rho }\partial _{\sigma }A_{\mu }\right)  
\nonumber \\
&&+A_{\rho }\partial _{\sigma }\left( \overline{c}c\right) +\overline{c}%
\partial _{\rho }c\partial _{\sigma }A_{\mu }-\overline{c}\partial _{\rho
}A_{\mu }\partial _{\sigma }c  \nonumber \\
&&+i\left( \partial _{\rho }b\partial _{\sigma }\partial ^{\mu }A_{\mu
}-\partial _{\rho }\overline{c}\partial _{\sigma }\partial ^{2}c\right)
{\LARGE ]}  \nonumber \\
&&-\partial ^{\mu }\theta ^{\rho \sigma }\left( x\right) b\left( A_{\sigma
}F_{\rho \mu }-A_{\rho }\partial _{\sigma }A_{\mu }\right) .  
\end{eqnarray}

This action corresponds to a highly nonlinear gauge.

The external field contribution is given by

\bigskip

\begin{equation}
\widehat{S}_{ext}=\int d^{4}x\quad \left( \widehat{A}^{\ast \mu }\star 
\widehat{s}\widehat{A}_{\mu }+\widehat{c}^{\ast }\star \widehat{s}\widehat{c}%
\right) \left( x\right) ,  
\end{equation}

\bigskip

where $\widehat{A}^{\ast }$, $\widehat{c}^{\ast }$ are external fields
(called antifields in the Batalin-Vilkovisky formalism) and play the role of
sources for the BRST-variations of the fields $\widehat{A}$, $\widehat{c}$.

The $\widehat{c}$ and $\widehat{\overline{c}}$ play quite asymmetric roles,
they cannot be related by Hermitian conjugation. The anti-BRST
transformations \cite{curci, ojima, baulieu} are given by

\bigskip

\begin{eqnarray}
\widehat{\overline{s}}\widehat{A}_{\mu } &=&\partial _{\mu }\widehat{%
\overline{c}}-\theta ^{\rho \sigma }\left( x\right) \left( \partial _{\rho }%
\overline{c}\partial _{\sigma }A_{\mu }-\partial _{\rho }A_{\mu }\partial
_{\sigma }\overline{c}\right) +o\left( \eta ^{2}\right) ,  \nonumber \\
\widehat{\overline{s}}\widehat{\psi } &=&i\widehat{\overline{c}}\widehat{%
\psi }-\theta ^{\rho \sigma }\left( x\right) \partial _{\rho }\overline{c}%
\partial _{\sigma }\psi +o\left( \eta ^{2}\right) ,  \nonumber \\
\widehat{\overline{s}}\widehat{\overline{\psi }} &=&-i\widehat{\overline{%
\psi }}\widehat{\overline{c}}+\theta ^{\rho }\left( x\right) \partial _{\rho
}\overline{\psi }\partial _{\sigma }\overline{c}+o\left( \eta ^{2}\right) ,
\nonumber \\
\widehat{\overline{s}}\widehat{F}_{\mu \nu } &=&-\theta ^{\rho \sigma
}\left( x\right) \left( \partial _{\rho }\overline{c}\partial _{\sigma
}F_{\mu \nu }-\partial _{\rho }F_{\mu \nu }\partial _{\sigma }\overline{c}%
\right) +o\left( \eta ^{2}\right) ,  \nonumber \\
\widehat{\overline{s}}\widehat{\overline{c}} &=&0,\quad \widehat{\overline{s}%
}\widehat{c}=-b,\quad \widehat{\overline{s}}\widehat{b}=0. 
\end{eqnarray}

\bigskip

Here $\widehat{\overline{s}}$ is the quantum anti-BRST operator. The
complete tree-level action is given by:

\bigskip

\begin{equation}
\Sigma \left( \widehat{A}_{\mu },\widehat{c},\widehat{\overline{c}},\widehat{%
b},\widehat{A}_{\mu }^{\ast },\widehat{c}^{\ast }\right) =\widehat{S}+%
\widehat{S}_{gf}+\widehat{S}_{ext}. 
\end{equation}

\bigskip

\section{Concluding Remarks}

We have defined a $q$-deformed QED at the classical level. Like in
noncommutative geometry \cite{bichl} we have found that the $q$-deformed QED
action contains non-renormalizable vertices of dimension six. It is
worthwhile to study the quantization of the $\eta $-expanded noncommutative $%
U\left( 1\right) $ Yang-Mills action and $q$-deformed BF Yang-Mills theory
\cite{mesref1} . We postpone these investigations to a future work.

Finally, we can claim, \textit{in bona fides}, that the method developed in
this letter can be applied to various quantum $q$-deformed models as well as
to $h$-deformed model (the so-called Jordanian models) \cite{mesref4,mesref5}%
.

\bigskip

\textbf{Acknowledgments}

I am very grateful to W. R\"{u}hl for reading the manuscript and the useful
discussions.

\end{document}